\documentclass[sigconf]{acmart}

\renewcommand\footnotetextcopyrightpermission[1]{} 
\usepackage{booktabs} 

\usepackage{natbib} 
\usepackage{silence}
\usepackage{xcolor}

\usepackage{soul}  
\setstcolor{red} 

\usepackage[textsize=scriptsize,backgroundcolor=green]{todonotes}

\definecolor{mycolor}{rgb}{0.037, 0.094, 0.105}

\setcopyright{rightsretained}

\begin{document}
\title{CobWeb: A Research Prototype for Exploring User Bias in Political Fact-Checking }

\author{Anubrata Das}
\affiliation{%
  \institution{School of Information\\University of Texas at Austin}
}
\email{anubrata@utexas.edu}

\author{Kunjan Mehta}
\affiliation{%
  \institution{School of Information\\University of Texas at Austin}
}
\email{kunjanmehta@utexas.edu}

\author{Matthew Lease}
\affiliation{%
  \institution{School of Information\\University of Texas at Austin}}
\email{ml@utexas.edu}


\begin{abstract}

The effect of user bias in fact-checking
has not been explored extensively from a user-experience perspective. 
We estimate the user bias as a function of the user's perceived reputation of the news sources (e.g., a user with liberal beliefs may tend to trust liberal sources). We build an interface to communicate the role of estimated user bias in the context of a fact-checking task. We also explore the utility of helping users visualize their detected level of bias. 80\% of the users of our system find that the presence of an indicator for user bias is useful in judging the veracity of a political claim.


\end{abstract}

%



\maketitle

\section{Introduction} \label{Introduction}

\begin{figure*}[t!]
\includegraphics[scale=0.35]{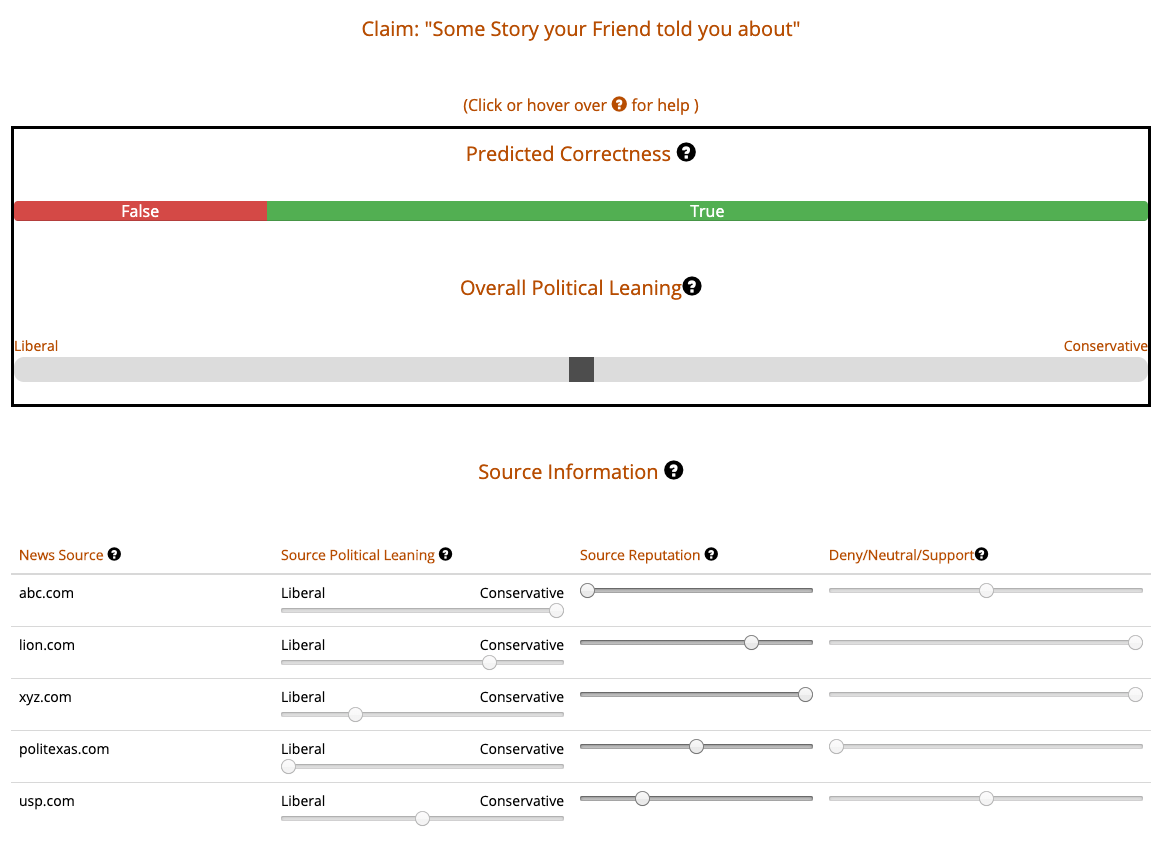}
\caption{The CoBWeb interface with claim and bias scores. Our interface extends  \citet{nguyen2018believe}'s design and prototype. The sliders indicating \textit{Overall Political Leaning} and \textit{Source Reputation} are enabled or disabled based on the experiments. In experiment 1 (Section \ref{exp1}), we familiarize the users with the interface by providing them with a set of movable sliders indicating the source reputation. We estimate the user's overall political leaning based on their interaction. In experiment 2 (Section \ref{exp2}), we enable the slider for overall political leaning for the users to change. Any change in the overall political leaning is reflected in the reputation score of the sources. In experiment 3 (Section \ref{exp3}), we provide the user with the interface shown in the image. This experiment is designed to understand if communicating overall political leaning helps the users in assessing the credibility of a claim. The elements used in each experiment are described in Table \ref{tab:components}.}
\label{System}
\end{figure*}
Recommender systems have the potential to reinforce existing user biases. This narrowing effect of recommender systems is called the \textit{Filter Bubble} \cite{pariser2011filter}. Users consume tailored content in systems such as e-commerce platforms, search engines, and social media. Recommender systems are capable of recording the users' responses and in real-time adjusting which recommendations are made. These systems regularly filter which information is exposed to the user. Filter bubbles are especially problematic due to the pervasive use of these recommendation systems.

In the context of news consumption, filter bubbles can impact users' social and political opinions. Social media and search engines have become a medium for news consumption, and recommender systems are an integral part of both social media and search engines. The news delivered to users also can be tailored to further reinforce existing beliefs. Similarly, the term \textit{Echo Chamber} \cite{jamieson2008echo} is used to describe the phenomenon when a group of people repeat each other's views without thoroughly analyzing and questioning them. Echo chambers play a significant role in reinforcing user biases both in online and offline news consumption.

Because filter bubbles create ideal echo-chambers, they are even more problematic in the context of spreading misinformation. Social media and search engines make it easy for users to access information, but the entry barrier for information creation and propagation is also low. It is important to note that spreading of misinformation is influenced by users' biases. Increasing personalization in the online media system can reinforce these biases. Different research communities have addressed the issue of fact-checking without taking the effect of user bias into account. We attempt to make users aware of the Filter Bubble in the context of verifying misinformation in political news \cite{iyyer2014political, popat2018credeye}.


\citet{nguyen2018believe, nguyen2018interpretable} introduce back-end algorithms and front-end interface design to support human fact-checking, demonstrated in a functional prototype\footnote{\url{http://fcweb.pythonanywhere.com}}. To enable transparency of the underlying prediction model, their prototype allows the users to dynamically modify the reputation of the news source and the stance of the article. Through interaction, users can better understand how such information can be aggregated to better estimate the veracity of online claims.

In this work, we extend \citet{nguyen2018interpretable}'s interface design by introducing an interactive slider for depicting the user's estimated political bias. Given the user's input on the reputation of the news sources, and bias values of these news sources, we estimate the user's bias. Also, the user's bias setting can be explicitly altered via user controls, enabling a user to observe the resulting modification in the reputation of the sources. We refer to our website as \textbf{Co}mmunication of \textbf{B}ias (in fact-checking) \textbf{Web}app (\textbf{CoBWeb})\footnote{\url{http://anubrata.pythonanywhere.com/biastask2/}}.

The rest of the article is organized as follows. We discuss the research questions and the contributions of this project in Subsection \ref{RQ} and \ref{Contribution}, respectively. In Section \ref{Related Work} we discuss the related works in this area. Methodologies used and experiments to be performed are discussed in Section \ref{Methods}. We discuss experiments and results in Section \ref{Frontend} and conclude the article in Section \ref{Conclusion}.

\subsection{Research Questions} \label{RQ}
With the increasing amount of interaction data used in social media platforms and search engines, users may see less and less information that challenges their beliefs. Often, it becomes difficult for users to realize that they are in a filter bubble. Behavioral echo chambers hinder users from making an informed judgment about the veracity of fake claims, especially in case of political news. 
We explore whether users' perceived reputation of news sources is an indicator of their political leaning i.e. bias.  By displaying the users' political leaning, we aim to provide users with control and transparency so that they can remain objective while judging the veracity of a claim.

As users become more aware of their own potential biases, they may also become more open to other perspectives and better assess the veracity of online information. We estimate user bias in CoBWeb and display it as an interactive slider widget. 
By using the slider to alter the model's estimate of user bias, the user can explore how their own potential bias may color their perceived reputation of news sources, as well as reflect on and assess whether the estimated user bias correctly reflects his/her own mental model.

    

\begin{description}

\item[RQ1]  How can we assist users in identifying their own biases in a political claim checking scenario?
\item[RQ2] In supporting users with fact-checking, how useful is it to show users an estimate of their own biases? 

\end{description}


\subsection{Contribution} \label{Contribution}
To the best of our knowledge, no prior work combines estimation and communication of user's political bias with fact-checking. As online echo chambers are the perfect space for spreading fake news, it is essential to study both of these phenomena together. In this work, our contributions are two-fold. First, we design and develop a user interface for communicating biases when users are engaged in political fact-checking. Second, we measure the effectiveness of our design in communicating bias and in verifying claims. We develop a prototype to enable users to view their estimated bias. Using our prototype, users can explore how alternative biases would change the predicted veracity of a given claim based on relevant news sources used in prediction. 

\section{Related Work} \label{Related Work}
Public opinion can be shaped by manipulating online media such as search engines. 
The effect of such manipulation can have a direct impact on the real world, such as changing a user's political opinion. \citet{epstein2015search} show that, depending on search engine results, voting preferences of as many as 20\% of the users could completely change, which might affect the outcome of an election. Communicating user bias and enabling a user to control the bias inferred and modeled by the system thus has the potential to support more informed decision making.

\citet{liao2013beyond} argue that despite the presence of competing views, users have a natural tendency to make choices that reinforce their own beliefs. They also suggest that biases can be controlled by presenting users with diverse perspectives, particularly when they are actively seeking information. Similarly, \citet{resnick2013bursting} present two strategies to avoid Filter Bubbles. The first strategy is to include diversified content in systems such as search engines and social media. The second strategy is to motivate the users to seek information actively. In this work, we provide users with indicators and controls for their biases, which in turn might motivate them toward more active information seeking.

\citet{nagulendra2014understanding} study the Filter Bubble in a social network with an interactive visualization prototype. They show that visualizations help users to become aware of the bubble. Based on the study by \citeauthor{nagulendra2014understanding}, we argue that it is essential to draw users' attention to the tangible effect of Filter Bubbles. In contrast, \citet{zuiderveen2016should} show that in the specific case of news sites, there is no empirical evidence that Filter Bubbles exist in algorithm-generated personalization. However, \citet{shearer2017news} show that around 67\% of adults in the United States get their news from social media instead of news sites. \citet{flaxman2016filter} show that there is an increase in the gap between different political ideologies in users who consume news through social media. Consequently, we argue that despite the lack of substantial evidence of filter bubbles in news sites, there is a need for studying the effects of the filter bubble in the context of social media and search engines.

A division of opinion amongst scholars exists regarding the impact of the echo chamber phenomenon in online news consumption. Some scholars argue that the phenomenon exists due to personalization, while some argue that it is an effect of a user's inherent tendency to reinforce his/her own beliefs. \citet{garrett2009echo} studies the phenomenon of echo-chambers in online news consumption. He argues that when users read news online, the difference between a user's exposure to opinion-reinforcing content and opinion-challenging content is not statistically significant. In contrast, \citet{quattrociocchi2016echo} present empirical evidence that users tend to reinforce their bias by ignoring arguments that contradict their beliefs. On the other hand, \citet{hannak2013measuring} show that in the case of search engines there is a significant difference in search results in the presence of a recommendation system.

\citet{lease2018fact} argues that online systems used for information gathering can be harmful for users if misinformation is provided in the context of fake news. He also discusses the importance of fact-checking in online information seeking and poses a set of related research questions for information retrieval. Our work builds on one of the research questions: how can we provide users with diverse information without catering to their political bias. Different applications address the problem of political bias on social media, such as \textit{Politecho}\footnote{\url{http://politecho.org/}} (not to be confused with Politico\footnote{\url{https://www.politico.com/}}) and \textit{Rbutr}\footnote{\url{http://rbutr.com}}. \textit{Politecho} is a Chrome extension that allows users to see where their Facebook friends are in the political spectrum and predict the political biases of Politecho users. \textit{Rbutr} is another Chrome extension that enables users to view sources that have both supported and rebutted the topic of a given web page. In our study, we aim to make these echo-chambers explicit in our user interface by communicating the user bias.

Prior work highlights that to address the problem of fact-checking, it is important to provide a user experience along with giving evidence regarding a claim \cite{bountouridis2018explaining, stab2018argumentext, popat2018credeye, nguyen2018believe, Chen-et-al:2019:NAACL}. \citet{bountouridis2018explaining} show that presentation of the evidence regarding a claim plays a crucial role in making a fact-checking system useful. \citet{popat2018credeye} propose a fact-checking system that uses language style, stance, and source reputation to predict the validity of a claim. They also show that highlighting an important section from the evidence to provide an explanation is more effective than just providing the evidence. \citet{nguyen2018believe} develop a similar system with an interpretable model for fact-checking. \citet{Chen-et-al:2019:NAACL} show that presenting the user with different perspective helps user in making better judgment towards a claim. Across these works, we see that user experience design play an important role in enabling effective communication with users. In this project, we create an interpretable user experience that helps users to better recognize and understand the effect of their own biases. Previous study in health information retrieval \cite{lau2009can} show that incorporating debiasing strategies in user interface design can influence a user's ability to interpret information in a more fair manner. 

\citet{zuiderveen2016should} mention that if users don't know that their search-engine webpage is personalized, they may not realize that different people are shown different results. The authors argue that transparency alone cannot solve the problem of filter bubbles; nevertheless, it is an important factor. The study also mentions the difference between self-selected personalization and pre-selected personalization of news. They find that self-selected personalization has an effect on influencing political attitudes, but this effect is very low. Also, \citet{zuiderveen2016should} argue that in the present technological landscape, pre-selected personalization cannot cause filter bubbles. This idea is contradictory to the caution indicated by \citet{liao2013beyond} about the negative effects of filter bubbles. This contradiction could rise because most studies are done on the US two-party system, whereas the study by \citet{zuiderveen2016should} refers to a multi-party democratic system.

In the political space, \citet{fossen2014s} challenge the existing Voting Advice Applications (VAAs) and aim to improve these applications' decisions by making them more transparent to the user. They argue that when a user is matched with a party using a specific set of criteria, it limits the perception of the user for that particular party. This effect on the user's mind is contrary to the aim of VAAs. Connecting this with our study, our aim is to make users' biases more transparent to them and potentially, help them perceive political information in a more neutral way.

\citet{tromp2011design} mention the recent attempts in design methodologies to deliberately change user behavior. The authors acknowledge that these methodologies have been rarely studied from effectiveness and ethical perspective. Their study is based on the assumption that the user is more or less receptive to behavior change. They conclude that the choice of design strategy for influencing behavior should be based on the kind of the intended user experience. They also provide specific design strategies for many scenarios, which we will incorporate in the design of our interface. We want our design to help users achieve a more neutral and balanced assessment of political claims.

Our prototype, CobWeb, with its movable sliders lies in the scope of Interactive Information Retrieval (IIR) systems. \citet{kelly2009methods} summarize the existing evaluation work in the field of IIR and offer suggestions for conducting user studies for IIR systems. Similarly,  \citet{cramer2008effects} take inputs from users and provide art recommendations based on the user's preferences. They have three experiments which have non-transparency, transparency and visibility of confidence levels in the interfaces, respectively. They find that transparency of the recommendation system improved acceptance of the recommendations and was better understood by the participants, but had no effect on the trust of the system. They also found that showing the confidence level of the system does not have an influence on the trust on the system. This is similar to the observation by \citet{nguyen2018interpretable}, in which the result of the perceived veracity of claims was not influenced even when the participants were aware that the system is fallible. \citet{sinha2002role} also find that transparency improves confidence on the recommendations. They show that users liked viewing the reasoning behind the recommendations for new items as well as already liked items. Hence, it is observed that transparency is a positive phenomenon and is preferred by the users.

\citet{pu2011user} introduce a framework called ResQue for doing user-centric evaluations for recommendation systems. They come up with thirty-two questions to measure user experience of a recommendation system based on metrics for perceived usefulness, perceived ease of use, control and transparency, etc. The short version of these questions is adopted for the study of our interface. \citeauthor{davis1989perceived} \cite{davis1989perceived} gives additional specifications on how to use the before-mentioned metrics of user experience (UX), which are also incorporated in our user studies.

\section{Methodology} \label{Methods}

In this section, we discuss the methodologies employed to estimate user bias. The proposed methodology for this project is two-fold, i.e., algorithmic component to estimate user bias, and user studies for evaluation. We estimate the user bias using bias of the news sources and how a user modifies the reputation of a news source in the context of a claim. 

The method by \citet{nguyen2018interpretable} is used to predict the reputation of a source given a claim. Based on a claim, a search is done on a commercial search engine and titles of articles which are relevant are retrieved and compared with the claim to understand whether it supports, refutes or is neutral to the claim. The system by \citet{nguyen2018interpretable} also uses a veracity classifier that models reputation score of a particular news source. 

Our prototype, CobWeb, extends the research prototype of \citet{nguyen2018believe} by introducing an additional slider to indicate estimated user bias. We also communicate the (estimated) political leaning of each source using a slider. If the user adjusts the reputation for a given source, as in \citet{nguyen2018believe}'s work, the predicted veracity of a claim is updated in real-time. In our work, the estimate for user bias is also updated as a function of how the user adjusts reputations assigned to sources (given the estimated political leanings of each source). In the other direction, any change in the estimated user bias also propagates to adjust the  reputation assigned by the model to news sources. The bias slider starts at a neutral point and ranges from ``Liberal'' to ``Conservative'' on the left and the right side of the slider, respectively. We show the shift in the overall bias as a relative scale for users to understand their position in the bias spectrum. 



\begin{table}[t!]
	\caption{An example of News Sources and their Bias scores}
    \label{tab:sourcebias}
	\begin{tabular}{|l|r|}
    	\hline
			\textbf{News Source} & \textbf{Bias} \\ \hline
			abc.com     & +1 \\ \hline
			lion.com    & +0.5  \\ \hline
			xyz.com     & -0.5  \\ \hline
			texas.times & -1 \\ \hline
			uta.edu     & 0    \\ \hline
	\end{tabular}
\end{table}

\subsection{\textbf{Estimating User Bias}} \label{User Bias}
We allow users to interact with the system in two distinct ways. A user can change both the estimated user bias and the source reputations for a particular claim. Once the bias score is changed, a user cannot change the reputation score, and vice-versa. Given this constraint, there are two mathematical relationships we need to establish. They are described below. 
\subsubsection{Change in Reputation Score}
When a user changes the reputation score of a particular source (for example, abc.com from Table \ref{tab:sourcebias}) from $R_1$ to $R_2$, the change in reputation score is represented by $\delta R = R_2 - R_1$. The change in user's overall bias is represented by $\delta \beta$. We calculate $\delta \beta$ using the formula below. 
\begin{equation}
\delta \beta = \delta R * Source_{bias}
\end{equation}
Once a user modifies the reputation score of a reputation slider, we shift the bias-slider by $\delta \beta$. For the purpose of interface design, in this study, we assign arbitrary bias scores ($Source_{bias}$) and reputation scores to imaginary sources, as shown in Table \ref{tab:sourcebias}. It is important to note that, if sources are neutral (for example, uta.edu from Table \ref{tab:sourcebias}), the change in reputation score does not necessarily change the estimated user bias score. 

\subsubsection{Change in Bias Score}
Our assumption is when the claims are checked using unmodified and model-estimated source reputation and stance, the user does not impose his/her bias on the system. Hence, the initial position of the user bias slider is thus always set to 0.  The user can move the user bias slider in either direction. Suppose the magnitude of the change is $\delta \beta$ and the sign denote the direction (-ve denotes left, and +ve denotes right). For each source, the change in reputations is calculated using the following.

\begin{equation}
    \delta R_{source} \propto \delta \beta
    \implies \delta R_{source} = Bias_{Source} * \delta \beta
\end{equation}

We use $\delta R_{source}$ to change the position of the reputation slider. Using the methodology discussed above, we aim to provide the user with an interaction method to enable them to understand and control their biases. Our goal is to investigate research questions RQ1 and RQ2. 

In order to evaluate our methodology, we perform a set of experiments with users to validate the usefulness and ease of use of use of our application. The details of the user study are discussed in Section \ref{Frontend}. Although political bias can be subjective and can vary in different contexts, we focus on the American political spectrum. 
%

\subsection{Design Decisions} \label{Design}

Developing an interface to communicate user bias raised several challenges and design questions. Note that the design questions are different from the research questions mentioned before in section \ref{RQ}. 
\begin{enumerate}
\item Which design element is effective for communicating bias?
\item How can we design the interaction between the reputation of news sources and the overall estimated bias of the user? 
\item How can we make the interaction interpretable for the user? 
\end{enumerate}

Another challenge was to evaluate the effectiveness of the interaction provided. We have created a step-by-step approach for evaluation. First, we familiarize the user with our interface, and then we evaluate their understandability of the system. Experimental design is described in details in Section \ref{Frontend}.

\subsubsection{Design Elements.}\citet{nguyen2018believe} use sliders as the main design elements to communicate the reputation of news sources and stance of news articles regarding a claim. They also use a movable bar to indicate the veracity of a claim. Drawing from their design, we chose the slider as our primary design element as well. Use of sliders provides us a thematic consistency across our application. 
In order to leverage the codebase provided by \citet{nguyen2018believe} we have used Flask\footnote{\url{https://flask.pocoo.org}}. 
%
We used a five point slider to limit user choices in changing the value of source reputation and overall bias. 
Since, the word \textit{bias} has a negative connotation to it, we have used the words \textit{overall political leaning} to indicate user bias. 
It is important to note that, we have used fictional data in the user study to prevent the influence of users' previous knowledge about actual news sources.
\subsubsection{Interpretability.} We aimed to enable users to understand our application's rationale for estimating their bias. As we have described the bias estimation in Section \ref{Methods}, the essential component for estimating the overall political leaning is the leaning of the news source. To communicate the news source biases we have used sliders that range from -1 to +1. The difference between the reputation sliders and the bias sliders are that the latter ones are not movable by the user. To show that the source bias sliders denote fixed values, we have grayed them out\footnote{\url{https://en.wikipedia.org/wiki/Grayed_out}}. To denote the range of the bias sliders, we have also included the words ``Liberal'' and ``Conservative'' on the left and the right side of each of these sliders. We have also explored other design elements such as using only words to denote the bias of the news source, but the information conveyed to the users was less granular. It was difficult to understand why changing the reputation of different sources by the same value would cause a varying degree of movement in the user bias slider.
\section{User Study} \label{Frontend}
The user study is divided into three experiments described later in this section, each consisting of several tasks. All participants perform each experiment. The survey experiments start with instructions on how to complete the experiment, and the definitions used in the survey. Participants are asked to use a laptop or a desktop computer, because our prototype is less mobile-friendly.

\subsection{\textbf{Website integration with Qualtrics}}
We display task instructions and collect responses on Qualtrics\footnote{\url{https://www.qualtrics.com/}}. We use Qualtrics because HTML, CSS and JavaScript codes can be easily integrated into it. Our prototype, CoBWeb, is integrated with Qualtrics by adding inline frame (iframe) blocks. We tested the integration and flow of the survey by doing a pilot study.

\subsection{\textbf{Recruitment and Respondents}}
We sent the survey link to potential participants, using University mailing lists, and got 10 responses. We did not collect participant demographics but believe that most of the participants are graduate students. 

\subsection{\textbf{Experiment 1}} \label{exp1}
This experiment has 10 tasks to assess whether participants understand the user bias slider   based on changing source reputations.

\subsubsection{Website elements}
At the top, we show the user bias (i.e., overall political leaning) slider. The overall political leaning slider is immovable for this experiment, to avoid confusion for the participants. Below the slider are three columns for news sources, political leaning of the news sources (i.e., source bias) and reputation of the news sources. We use a stripped down version of Figure \ref{System} for this experiment by removing the stance sliders and predicted correctness slider. In table \ref{tab:components} we list the components that appear in the interface for this experiment. 


\subsubsection{Procedure}
To accustom the participants to the system, the first five tasks ask the participant to change the source reputation in a specific direction and then observe changes in the inferred user's bias. Next, they answer a question on a 5 point Likert scale about the degree to which they agree with the system's estimate of inferred user's bias. An example of such a task is as follows:

\textit{Imagine you have a Neutral perspective for politics and believe that abc.com should have a higher reputation. Change the source reputation for the news source and observe the change in the overall political leaning.}

We ask each participant five such questions for  five different news sources to learn whether the participants understand the system. To confirm the their understanding of the system, we ask another five questions. The participants change the reputations of five sources on the website  and estimate the resulting inferred user's bias in each case. To avoid influencing participants' predictions, we keep the algorithm's prediction of user bias stationary when source reputations are modified. We collect the responses on a five-point scale ranging from ``Extreme Liberal'' to ``Extreme Conservative''. Next, we compare participants' responses with the answers generated by our algorithm.

\subsubsection{Results}
For the initial five tasks where we ask the participants the degree to which they agree with the system, we find that 9 out of 10 participants have selected ``Agree'' or ``Strongly Agree'' on the Likert scale. Hence, participants tend to agree with the algorithm's prediction of user bias.

In the next five tasks, which ask the users to predict the user bias, we find that 6 out of 10 participants have gotten at least 60\% of the answers correct. It seems that participants tend to trust the system, and hence they agree with it even when they do not completely understand the way it works.

For the task which has a neutral news source (neither Liberal nor Conservative), 9 out of 10 participants have predicted the user bias correctly. Hence, it seems that the predicted user bias is more understandable when the news sources are neutral rather than Liberal or Conservative.

Conversely, when the source bias is not neutral, and source reputation is set to 0.5 (0 being the lowest reputation, and 1 being the highest reputation), only 3 out of 10 participants have gotten the answer right. More than half of the participants (6 out of 10) believe that the user bias slider should point to neutral. This result indicates that the participants are confused between the terms ``source reputation'' and ``source bias''.

\subsection{\textbf{Experiment 2}} \label{exp2}
This experiment investigates whether change in source reputation is understood by the participants when the user bias is modified.

\subsubsection{Website elements}
The website looks the same as in Experiment 1. The only difference is that the user bias slider is movable in this experiment.

\subsubsection{Procedure}
There are two tasks in this experiment, the first task is designed to accustom the participants to the system and how it works. The task requires them to change the position of the user bias slider and observe changes in the source reputations. Next, the participants mark their degree of agreement with the algorithm's prediction of source reputations on a Likert scale. For example, when the position of the user bias slider is moved towards the ``Liberal'', it leads to an increase in the reputation scores for sources having a Liberal political leaning.

To know whether the participants have understood the above concept, the second task asks the participants to predict the source reputations when the user bias slider is moved in a given direction. To avoid influencing participants' predictions, we mask the algorithm's prediction by not reflecting the change caused by modification in position of the user bias slider.

\subsubsection{Results}
For the first task, only half of the participants agree with the algorithm's prediction of source reputations. In the second task, 6 out of 10 participants answered at least 60\% of the answers correct. Even though the participants only partially agree with the algorithm's prediction when it is explicitly asked, the participants' answers in the second task imply that they agree with the algorithm's prediction more than their prior belief. This degree of agreement also suggests that the estimate of source reputation is understandable to most of the participants.

\subsection{\textbf{Experiment 3}} \label{exp3}
This experiment assesses participants' accuracy to predict truthfulness of a political claim when user bias slider is present on the website.

\subsubsection{Website elements}
For this experiment, the website has all the elements as earlier with two additional elements: the algorithm's prediction of whether the claim is true and the stance of each news source. The stance shows whether a news source denies or supports a political claim.

\subsubsection{Procedure}
There are 5 tasks in this experiment. We ask participants to assume a claim like ``Clinton and Trump are friends on Facebook''. In each task, participants modify the source reputation of a news source and observe changes in system prediction of claim veracity and the inferred user's bias. Next, the participants assess claim veracity.

\subsubsection{Results}
We compare the participants' and the algorithm's predictions for the truthfulness of a given claim. 8 out of 10 participants have the same prediction as the algorithm's prediction. Hence, when the algorithm's predicted veracity and user bias is shown on the website, 80\% of the participants can correctly interpret the correctness of the claim.

\subsection{\textbf{Post-tasks questionnaire}}
\citet{davis1989perceived} defines perceived usefulness as ``the degree to which a person believes that using a particular system would enhance his/her job performance'' (p. 320). He introduces six questions to determine the perceived usefulness of a system. We adopt those questions and modify them in the perspective of communicating user's bias. Out of the six questions, we eliminate two questions which are related to speed and productivity, as these questions did not relate to our study's goal.

We calculate the average score of each participant and find that 80 percent of the participants believe that the user bias slider is a useful feature in predicting the truthfulness of a claim.






\section{Limitations}
This study assumes a simplified American political fact-checking context with the user bias and source bias values from Liberal to Conservative. It will be more complicated to apply this approach in domains having variables in more than two dimensions.

There are relatively few participants in this study and they are largely graduate students. In the future, the scale and diversity of participants should be further increased.

We have used synthetic data in the study to control independent variables. When users interact in the real world, they may have preset beliefs for certain news sources which may influence the way they interact with our system. They may not agree with the algorithm's prediction of user bias if it opposes their beliefs. Hence, the system needs to be tested with real news sources, and their real reputations and biases in order to have more conclusive results.

\begin{table*}[t!]
\centering
\resizebox{0.95\textwidth}{!}{
\begin{tabular}{llll}
\textbf{Experiment} & \textbf{Description}                                                                                                                                                     & \textbf{Components}                                                                       & \textbf{Type}       \\ \hline
1                   & Change in source reputation reflects overall political leaning of a user                                                                                                 & News Sources                                                                              & Static Text         \\
                    &                                                                                                                                                                          & Reputation of the sources                                                                 & Input               \\
                    &                                                                                                                                                                          & Political Leaning of the Sources                                                          & Constant            \\
                    &                                                                                                                                                                          & User's overall political leaning                                                          & Output              \\ \hline
2                   & Change in overall political leaning reflects on the reputation of the news sources                                                                                       & \begin{tabular}[c]{@{}l@{}}News Sources \\ Political Leaning of the Sources\end{tabular} & -- Same as above -- \\
                    &                                                                                                                                                                          & User's overall political leaning                                                          & Input               \\
                    &                                                                                                                                                                          & Reputation of the sources                                                                 & Output              \\ \hline
3                   & \begin{tabular}[c]{@{}l@{}}Combination of Overall Political Leaning and Predicted Correctness of a claim\\ helps users to assess the credibility of a claim\end{tabular} & \begin{tabular}[c]{@{}l@{}}News Sources \\ Political Leaning of the Sources\end{tabular}  & -- Same as above -- \\
                    &                                                                                                                                                                          & Stance                                                                                    & Constant            \\
                    &                                                                                                                                                                          & Reputation of the sources                                                                 & Input               \\
                    &                                                                                                                                                                          & User's overall political leaning                                                          & Output              \\
                    &                                                                                                                                                                          & Predicted correctness of a claim                                                          & Output              \\ \hline
\end{tabular}}
\caption{List of elements used in the interface for the experiments}
\label{tab:components}
\end{table*}

\section{Conclusion} \label{Conclusion}
We investigate the Filter Bubble problem using interaction design as a methodology. We provide a system design to estimate and communicate user bias in the context of fact-checking. Our design has potential to be generalized to other areas beyond politics, such as claims related to the effects of vaccination. The evaluation method for this work could also be more broadly applied to interaction-design for explainable machine learning.

The results of our study support that the participants tend to agree more with the algorithm's prediction of user bias than the algorithm's prediction of source reputation. In general, we can infer that the interplay between user bias and source reputation appears to be understood by the participants. This realization helps us understand how source reputation helps in calculating user bias (RQ1) and validates our approach of estimating the user  bias.

To predict claim truthfulness, participants tend to trust the algorithm's prediction even though we have mentioned in the study that the algorithm is typically correct only 70\% of the time. Overall, participants find the user bias slider useful in predicting the truthfulness of claims. Hence, the communication of user bias is understandable and useful for most participants.


In this work, we provide a design and web-application using hypothetical sources, data, and scenarios. In the future, we aim to implement the system using real data. We also plan to extend this work in areas such as sentiment-based bias. The findings and methods of this study can be employed to understand how the emotional state of a user affects their content consumption, especially in the fake news context.

\section*{Acknowledgements}
We thank Ricardo Baeza-Yates for the seed idea of making filter bubbles more visible to users. We also thank UT Austin's {\em Center for Media Engagement} (CME)\footnote{\url{https://moody.utexas.edu/research/center-media-engagement/}}, and particularly Natalie J.\ Stroud, for valuable conversations further informing our understanding of misinformation today in America's political landscape. This work is supported in part by the Micron Foundation, and by {\em Good Systems}\footnote{\url{https://bridgingbarriers.utexas.edu/good-systems/}}, a UT Austin Grand Challenge Initiative to develop socially-responsible AI technologies. We also thank the anonymous reviewers for giving insightful suggestions on this work. The statements made herein are solely the responsibility of the authors. 

\bibliography{references}


\end{document}